\documentclass[12pt,oneside,letter,rotating,leqno]{article}
\usepackage{array}
\usepackage[english]{babel}
\usepackage{amsmath}
\usepackage{amssymb}
\usepackage{amsthm}
\usepackage{latexsym}
\usepackage[pdftex]{graphicx,color}
\usepackage{makeidx}
\usepackage{harvard}
\usepackage{endnotes}
\usepackage[nolists,tablesfirst]{endfloat}
\usepackage{multirow}
\usepackage{rotating}
\usepackage{exercise}
\usepackage{lscape}
\usepackage{verbatim}
\usepackage[small,bf]{caption}
\usepackage{chngpage}
\usepackage{booktabs}
\usepackage{verbatim}

\usepackage{longtable}
\usepackage{booktabs}

\usepackage{answers}

\newcommand{\R}{\mathbb{R}}
\newcommand{\prob}[1]{\mathbb{P}\left[#1\right]}

{}

\newcommand{\A}{\mathbb{A}}
\newcommand{\LL}{\mathbb{L}}
\newcommand{\N}{\mathbb{N}}
\newcommand{\E}{\mathbb{E}}
\newcommand{\X}{\mathcal{X}}

\newcommand{\core}{\mathrm{core}}

\renewcommand{\baselinestretch}{1}

\begin{document}

\title{Diversification Preferences in the Theory of Choice}

\author{{\bf Enrico G. De Giorgi}\footnote{Department of Economics, School of Economics and Political Science, University of St. Gallen, Bodanstrasse 6, 9000 St. Gallen, Switzerland, Tel. +41 +71 2242430,
Fax. +41 +71 224 28 94, email:
enrico.degiorgi@unisg.ch.} \and {\bf
Ola Mahmoud}\footnote{Faculty of Mathematics and Statistics, School of Economics and Political Science, University of St. Gallen, Bodanstrasse 6, 9000 St. Gallen, Switzerland and Center for Risk Management Research, University of California, Berkeley, Evans Hall, CA 94720-3880, USA, email: ola.mahmoud@unisg.ch}
}
\date{\today}
\maketitle
 \linespread{1}
\begin{abstract}
Diversification represents the idea of choosing variety over uniformity. Within the theory of choice, desirability of diversification is axiomatized as preference for a convex combination of choices that are equivalently ranked. This corresponds to the notion of risk aversion when one assumes the von-Neumann-Morgenstern expected utility model, but the equivalence fails to hold in other models. This paper analyzes axiomatizations of the concept of diversification and their relationship to the related notions of risk aversion and convex preferences within different choice theoretic models. Implications of these notions on portfolio choice are discussed. We cover model-independent diversification preferences, preferences within models of choice under risk, including expected utility theory and the more general rank-dependent expected utility theory, as well as models of choice under uncertainty axiomatized via Choquet expected utility theory. Remarks on interpretations of diversification preferences within models of behavioral choice are given in the conclusion.

\mbox{}
\end{abstract}
\renewcommand{\baselinestretch}{1}

\newtheorem{defi}{Definition}
\newtheorem{lemma}{Lemma}
\newtheorem{ass}{Assumption}
\newtheorem{prop}{Proposition}
\newtheorem{cor}{Corollary}
\newtheorem{thm}{Theorem}
\newtheorem{rem}{Remark}
\newtheorem{axiom}{Axiom}

\newpage

\section{Introduction}

\begin{quote}
{\it Another rule which may prove useful can be derived from our theory. This is the rule that it is advisable to divide goods which are exposed to some danger into several portions rather than to risk them all together.}
\end{quote}

\hspace{4in} -- Daniel Bernoulli, 1738 \\

The term \emph{diversification} conveys the idea of introducing variety to a set of objects. Conceptually, \citeasnoun{Bernoulli} may have been the first to appreciate the benefits of diversification in an economic context. In his fundamental 1738 article on the St. Petersburg paradox, he argues by example that risk averse investors will want to diversify.

In finance, diversification is perhaps the most important of investment principles. Here, it is roughly understood as the mitigation of overall portfolio risk by investing in a wide variety of assets. The seminal work of \citeasnoun{Markowitz1952} on portfolio theory laid the first mathematical foundations for what we understand today under investment diversification in finance. Markowitz's portfolio theory provides a crucial formalization of the link between the inseparable notions of diversification and risk; it postulates that an investor should maximize portfolio return while minimizing portfolio risk, given by the return variance. Hence, diversification in finance is equivalent to the reduction of overall risk (but not generally its elimination). Interestingly, assuming that markets are populated by investors as in \citeasnoun{Markowitz1952}, only non-diversifiable risk is priced at equilibrium, as shown by the well-known Capital Asset Pricing Model (CAPM) developed by \citeasnoun{Sharpe:1964}, \citeasnoun{Lintner:1965}, and \citeasnoun{Mossin:1966}. Briefly, under the CAPM, investors are only rewarded for non-diversifiable or systematic risk.

Diversification plays an equally important role in economic theory. The way in which an individual economic agent makes a decision in a given choice theoretic model forms the basis for how diversification is viewed. An economic agent who chooses to diversify is understood to prefer variety over similarity. Axiomatically, a preference relation $\succsim$ on a set of choices $\X$ exhibits \emph{preference for diversification} if for any $x_1, \dots, x_n \in \X$ and $\alpha_1, \dots, \alpha_n \in [0,1]$ for which $\sum_{i=1}^n \alpha_i = 1$,
$$ x_1 \sim \dots \sim x_n \Rightarrow \sum_{i=1}^n \alpha_i\, x_i \succsim x_j \quad \textrm{for all } j=1,\dots,n.$$
An individual will hence want to diversify among a collection of choices all of which are ranked equivalently.
This notion of diversification is equivalent to that of convexity of preferences, which states that  $\alpha \,x + (1-\alpha)\,y\succsim y$, for all $\alpha\in [0,1]$, if $x\succsim y$.

The most common example of diversification in the above choice theoretic sense is within the universe of asset markets, where an investor faces a choice amongst risky positions, such as equities, derivatives, portfolios, etc.  Such risky positions are usually modeled as random variables on some state space $\Omega$ under a given objective reference probability $\mathbb{P}$. Diversification across two equivalently ranked risky assets $x$ and $y$ is then expressed by the state-wise convex combination $\alpha\,x(\omega) + (1-\alpha)\,y(\omega)$ for $\mathbb{P}$-almost all $\omega\in\Omega$ and $\alpha\in[0,1]$. In this context, preference for diversification means that an investor would prefer to allocate a fraction $\alpha$ to asset $x$ and a fraction $1-\alpha$ to asset $y$ rather than fully invest in either one of the assets. Within traditional decision theory under risk, where preferences are formed over lotteries, that is probability measures $p\,:\,Z\to[0,1]$ over a set of prizes $Z$, diversification has yet another interpretation.  Here, a convex combination $\alpha\,p + (1-\alpha)\,q$ of equally desirable lotteries $p$ and $q$ is defined by taking the convex combination of each prize $z$ separately, that is, $(\alpha\, p + (1-\alpha)\,q)(z) = \alpha\,p(z) + (1-\alpha)\,q(z)$. This convex combination of lotteries can be interpreted as some additional randomization, since it corresponds to the sampling of either $p$ or $q$ depending on the outcome of a binary lottery with probability $\alpha$ or $1-\alpha$. 

The concepts of diversification and risk aversion are closely intertwined. In axiomatic choice theory, risk aversion is roughly the preference for a certain outcome with a possibly lower payoff over an uncertain outcome with equal or higher expected value. More precisely, a decision maker is said to be risk averse if the expected value of a random variable with certainty is preferred to the random variable itself.
 Informally, one might say that the goal behind introducing variety through diversification is the reduction of ``risk" or ``uncertainty", and so one might identify a diversifying decision maker with a risk averse one. This is indeed the case in expected utility theory (EUT), where risk aversion and preference for diversification are exactly captured by the concavity of the utility function. However, this equivalence fails to hold in other models of choice.

Even though the desirability for diversification is a cornerstone of a broad range of portfolio choice models in finance and economics, the precise formal definition differs from model to model. Analogously, the way in which the notion of diversification is interpreted and implemented in the investment management community varies greatly. Diversifying strategies thus span a vast range, both in theory and in practice, from the classical approaches of Markowitz's variance minimization and von-Neumann-Morgenstern's expected utility maximization, to the more naive approaches of equal weighting or increasing the number of assets.

Diversification, in its essence, can be regarded as a choice heuristic that comes in different forms.
This article provides the first comprehensive overview of the various existing formalizations of the notion of diversification from a choice theoretic perspective. Different axiomatizations of the concept of diversification and their relationship to related notions of risk aversion are reviewed within some of the most common decision theoretic frameworks. Motivated by Bernoulli's realization of its financial and economic benefits, we discuss the implications of each of the major definitions of diversification on portfolio choice.

We start by setting up the theoretical choice theoretic framework in Section \ref{sec:theoretical setup}. Section \ref{section:rudiments} examines various choice theoretic axiomatizations of the concept of diversification and their relationship to convexity of preferences and concavity of utility. Given the intrinsic link between risk aversion and diversification, Section \ref{section:risk_aversion} reviews common definitions of risk aversion, including weak, strong and monotone risk aversion, and their inter-relationship. Section \ref{section:model_ind} studies the connection between diversification preferences, convex preferences and risk aversion under no particular model assumptions. Section \ref{section:eut} reviews the classic results within the framework of expected utility theory, where all definitions of diversification preferences as well as all notions of risk aversion coincide with the concavity of the von-Neumann-Morgenstern utility representation. Section \ref{section:rdeu} considers the more general rank-dependent expected utility model of \citeasnoun{Quiggin1982}, where the equivalence between weak and strong risk aversion does not carry over from the expected utility model. Similarly, the correspondence between risk aversion and preference for diversification does not hold any longer. Section \ref{section:ceu} extends the discussion from models of choice under risk to models of choice under uncertainty. This covers decision models where there is no given objective probability distribution on the set of states of the world, the axiomatization of which is given by Choquet expected utility theory. Notions of diversification and uncertainty aversion, rather than risk aversion, are discussed within such models of expected utility under a non-additive subjective probability measure. Under uncertainty, the phenomenon of \emph{ambiguity aversion} roughly captures the preference for known risks over unknown risks. In Section \ref{section:ambiguity}, we recall the most prominent definitions of ambiguity aversion, link them to the notion of diversification and discuss the implications on portfolio choice. Section \ref{section:conclusion} concludes with remarks on interpretations of diversification preferences within models of behavioral choice.

\section{Theoretical setup}\label{sec:theoretical setup}

We adopt the classical setup for risk assessment used in mathematical finance and portfolio choice. We consider a decision maker who chooses from the vector space $\X=\LL^\infty(\Omega,\mathcal{F},\mathbb{P})$ of essentially bounded real-valued random variables on a probability space $(\Omega, \mathcal{F}, \mathbb{P})$, where $\Omega$ is the set of states of nature, $\mathcal{F}$ is a $\sigma$-algebra of events, and $\mathbb{P}$ is a $\sigma$-additive probability measure on $(\Omega,\mathcal{F})$. Note that the decision maker is also able to form compound choices represented by the state-wise convex combination $\alpha\,x+(1-\alpha)\,y$ for $x,y\in\X$ and $\alpha\in[0,1]$, defined by $\alpha\,x(\omega) + (1-\alpha)\,y(\omega)$  for $\mathbb{P}$-almost all $\omega\in\Omega$. The space $\X$ is endowed with the order $x\ge y\Leftrightarrow x(\omega)\ge y(\omega)$ for $\mathbb{P}$-almost all $\omega\in\Omega$.
Our assumption that the outcome space is the set $\R$ of real numbers, which comes with an intrinsic ordering and mixing operations, enables a natural monetary interpretation of outcomes.\\


A weak preference relation on $\X$ is a binary relation $\succsim$ satisfying:
\begin{enumerate}
\item[(i)] \emph{Completeness}: For all $x,y \in \X$, $x \succsim y \lor y \succsim x$.
\item[(ii)] \emph{Transitivity}: For all $x,y,z \in \X$, $x\succsim y \land y \succsim z \Rightarrow x \succsim z$.
\end{enumerate}
Every weak preference relation $\succsim$ on $\X$ induces an indifference relation $\sim$ on $\X$ defined by $x \sim y \Leftrightarrow (x \succsim y)\land (y \succsim x)$. The corresponding strict preference relation $\succ$ on $\X$ is defined by $x \succ y \iff x\succsim y \land \neg(x \sim y)$. A numerical or utility representation of the preference relation $\succsim$ is a real-valued function $u \colon \X \to \R$ for which $x \succsim y \iff u(x) \geq u(y)$.\\

For $x\in\X$, $F_x$ denotes the cumulative distribution function of $x$, defined by $F_x(c) = \prob{x\leq c}$ for $c\in\R$, and $e(x)$ is the expectation of $x$, that is, $e(x) = \int c \ dF_x(c)$. For $c\in \R$, $\delta_c$ denotes the degenerated random variable with $\delta_c(\omega)=c$ for $\mathbb{P}$-almost all $\omega\in\Omega$. The \emph{certainty equivalent} of $x\in\X$ is the value $c(x)\in\R$ such that $x\sim\delta_{c(x)}$, i.e., $c(x)$ is the certain value which the decision maker views as equally desirable as a choice $x$ with uncertain outcome. The \emph{risk premium} $\pi(x)$ of $x\in\X$ is the amount by which the expected return of a choice $x\in\X$ must exceed the value of the guaranteed outcome in order to make the uncertain and certain choices equally attractive. Formally, it is defined as $\pi(x) = e(x)-c(x)$.

\paragraph{Monotonicity.}
Emulating the majority of frameworks of economic theory, it seems reasonable to assume that decision makers prefer more to less. In particular, in view of the monetary interpretation of the space $\X$, a natural assumption on the preference relation $\succsim$ is \emph{monotonicity}.
\begin{enumerate}
\item[(iii)] \emph{Monotonicity}: For all $ x,y\in\X,$ $x\geq y \implies  x \succsim y$.
\end{enumerate}
Monotonicity of preferences is equivalent to having a strictly increasing utility function $u$. Indeed, for $x\geq y$, we have $x\succsim y$ and thus $u(x)\geq u(y)$. Monotonicity of the utility function simply implies that an agent believes that ``more is better"; a larger outcome yields greater utility, and for risky bets the agent would prefer a bet which is first-order stochastically dominant over an alternative bet.

\paragraph{Continuity.} Continuity of preferences is often assumed for technical reasons, as it can be used as a sufficient condition for showing that preferences on infinite sets can have utility representations. It intuitively states that if $x\succ y$, then small deviations from $x$ or from $y$ will not reverse the ordering.
\begin{enumerate}
\item[(iv)] \emph{Continuity}: For every $x\succ y$, there exist neighborhoods $B_x,B_y\subseteq \X$ around $x$ and $y$, respectively, such that for every $x'\in B_x$ and $y'\in B_y$, $x'\succ y'$.
\end{enumerate}
Throughout this article, unless otherwise stated, we assume that preferences are both monotonic and continuous. Debreu's theorem \cite{Debreu1964} states that there exists a continuous monotonic utility representation $u$ of a monotonic and continuous preference relation $\succsim$.


\section{Rudiments of convexity, diversification, and risk}
\label{section:rudiments}

The notions of convex preferences, preferences for diversification, and risk are inherently linked, both conceptually and mathematically. We recall the formal definitions and their relationships.

\subsection{Convex preferences}

We begin with the mathematically more familiar concept of convexity. The notion of convexity of preferences inherently relates to the classic ideal of diversification, as introduced by \citeasnoun{Bernoulli}.
To be able to express convexity of a preference relation, one assumes a choice-mixing operation on $\X$ that allows agents to combine (that is diversify across) several choices. By combining two choices, the decision maker is ensured under convexity that he is never ``worse off" than the least preferred of these two choices.

\begin{defi}[Convex preferences]
A preference relation $\succsim$ on $\X$ is \emph{convex} if for all $x,y \in \X$ and for all $\alpha \in [0,1]$,
$$ x  \succsim y \implies \alpha \,x + (1-\alpha)\, y \succsim y . $$
\end{defi}

In mathematics and economic theory, convexity is an immensely useful property, particularly within optimization. The role it plays in the theory of choice leads to some convenient results, because of what it says about the corresponding utility representation. We recall some well-known properties of utility functions representing convex preferences.

\begin{prop}
A preference relation $\succsim$ on $\X$ is convex if and only if its utility representation $u \colon \X \to \R$ is quasi-concave.
\end{prop}

This means that convexity of preferences and quasi-concavity of utility are equivalent. A direct corollary to this result is that concavity of utility implies convexity of preferences. However, convex preferences may have numerical representations that are not concave.\footnote{To see this, suppose $u$ is a concave utility function representing a convex preference relation $\succsim$. Then if a function $f:\R\to\R$ is strictly increasing, the composite function $f\circ u$ is another utility representation of $\succsim$. However, for a given concave utility function $u$, one can relatively easily construct a strictly increasing function $f$ such that $f\circ u$ is not concave.}
More strongly even, some convex preferences can be constructed in a way that does not admit any concave utility representation.

\subsection{Diversification preferences}

An important property within the theory of choice is that of diversification. An economic agent who chooses to diversify is understood to prefer variety over similarity. Axiomatically, preference for diversification is formalized as follows; see \citeasnoun{Dekel1989}.

\begin{defi}[Preference for diversification]
A preference relation $\succsim$ exhibits \emph{preference for diversification} if for any $x_1, \dots, x_n \in \X$ and $\alpha_1, \dots, \alpha_n \in [0,1]$ for which $\sum_{i=1}^n \alpha_i = 1$,
$$ x_1 \sim \dots \sim x_n \implies \sum_{i=1}^n \alpha_i x_i \succsim x_j \quad \textrm{for all } j =1,\dots, n.$$
\end{defi}

This definition states that an individual will want to diversify among a collection of choices all of which are ranked equivalently. The most common example of such diversification is within the universe of asset markets, where an investor faces a choice amongst risky assets. We recall that this notion of diversification is, in our setup, equivalent to that of convexity of preferences.

\begin{prop}
A monotonic and continuous preference relation $\succsim$ on $\X$ is convex if and only if it exhibits preference for diversification.
\end{prop}

Various other definitions of diversification exist in the literature. \citeasnoun{Chateauneuf2002} introduce the stronger notion of \emph{sure diversification}. Roughly, sure diversification stipulates that if the decision maker is indifferent between a collection of choices and can attain certainty by a convex combination of these choices, he should prefer that certain combination to any of the uncertain choices used in the combination.

\begin{defi}[Preference for sure diversification]
A preference relation $\succsim$ exhibits \emph{preference for sure diversification} if for any $x_1, \dots, x_n \in\X$ and $\alpha_1, \dots, \alpha_n \geq 0$ satisfying $\sum_{i=1}^n \alpha_i = 1$, and $c,\beta\in\R$,
$$ \left[ x_1 \sim \cdots \sim x_n  \land \sum_{i=1}^n \alpha_i x_i = \beta\delta_c \right] \implies \beta\delta_c \succsim  x_i, \quad \forall i=1,\dots, n .$$
\end{defi}

\citeasnoun{Chateauneuf2007} introduce a weakening of the concept of preference for diversification, which is referred to as preference for \emph{strong diversification}. Preference for strong diversification means that the decision maker will want to diversify between two choices that are identically distributed.

\begin{defi}[Preference for strong diversification]
A preference relation $\succsim$ exhibits \emph{preference for strong diversification} if for all $x,y\in\X$ with $F_x=F_y$ and $\alpha\in[0,1]$, $\alpha x + (1-\alpha)y \succsim y$.
\end{defi}

Preference for diversification implies preference for sure diversification, but the converse does not hold. One can also show that preferences for strong diversification are equivalent to requiring that preferences respect second-order stochastic dominance (see \citeasnoun{Chateauneuf2007} for an outline of the proof). Moreover, \citeasnoun{Chateauneuf2007} also provide a counterexample showing that preferences for strong diversification do not imply convex preferences.

Yet another notion of diversification was introduced by \citeasnoun{Chateauneuf2002}, namely that of \emph{comonotone diversification}. Two random variables $x,y\in\X$ are \emph{comonotonic} if they yield the same ordering of the state space from best to worst; more formally if for every $\omega,\omega'\in S$, $(x(\omega)-x(\omega'))(y(\omega)- y(\omega'))\geq0$. Comonotonic diversification is defined as follows:

\begin{defi}[Comonotone diversification]
A decision maker exhibits preference for \emph{comonotone diversification} if for all comonotonic $x$ and $y$ for which $x\sim y$,
$$ \alpha x + (1-\alpha)y \succsim x $$
for all $\alpha \in (0,1)$.
\end{defi}

Comonotone diversification is essentially convexity of preferences restricted to comonotonic random variables, just as \citeasnoun{Schmeidler1989} restricted independence to comonotonic acts (see Section \ref{section:ceu} for a more detailed discussion of Schmeidler's model). Under this more restrictive type of diversification, any hedging in the sense of \citeasnoun{Wakker1990} is prohibited.

\subsection{Diversification and quasiconvex risk}

In mathematical finance, diversification is often understood to be a technique for reducing overall \emph{risk}, where, here, one may follow the Knightian \cite{Knight1921} identification of the notion of risk as \emph{measurable uncertainty}. In classical risk assessment within mathematical finance, uncertain portfolio outcomes over a fixed time horizon are represented as random variables on a probability space. A risk
measure maps each random variable to a real number summarizing the overall position in risky
assets of a portfolio.
In his seminal paper, \citeasnoun{Markowitz1952}, even though he proposed variance as a risk measure, emphasized the importance for a risk measure to encourage diversification.\footnote{``\emph{Diversification is both observed and sensible; a rule of behavior which does not imply the superiority of diversification must be rejected.}"} Over the past two decades, a number of academic efforts have more formally proposed properties that a risk measure should satisfy, for example the work of \citeasnoun{FollmerSchied2010} and \citeasnoun{FollmerSchied2011} who echo Markowitz in that a ``good'' risk measure needs to promote diversification. The key property is once again that of convexity, which, if satisfied, does not allow the diversified risk to exceed the individual standalone risks. It thus reflects the key principle of economics and finance, as well as the key normative statement in the theory of choice, namely that \emph{diversification should not increase risk}.

We review this diversification paradigm in the context of risk measures within the theory of choice. We emulate the formal setup of \citeasnoun{DrapeauKupper2013}, where the risk perception of choices is modeled via a binary relation $\hat\succsim$, the ``risk order", on $\X$ satisfying some appropriate normative properties. A risk order represents a decision maker's individual risk perception, where $x\hat\succsim y$ is interpreted as $x$ being \emph{less risky} than $y$. Risk measures are then quasiconvex monotone functions, which play the role of numerical representation of the risk order.

The two main properties of risk captured by a risk order are those of convexity and monotonicity. The convexity axiom reflects that diversification across two choices keeps the overall risk below the worse one; the monotonicity axiom states that the risk order is compatible with the vector preorder. A formal definition follows.

\begin{defi}[Risk order]
A \emph{risk order} on the set $\X$ is a reflexive weak preference relation satisfying the convexity\footnote{\citeasnoun{DrapeauKupper2013} refer to the convexity property as \emph{quasiconvexity}, which we believe is a mathematically more appropriate nomenclature. However, we stick to the more widely used \emph{convexity} terminology for consistency.} and monotonicity axioms.
\end{defi}

Numerical representations of risk orders inherit the two key properties of convexity and monotonicity of a decision maker's risk perception and are called risk measures.

\begin{defi}[Risk measure]
A real-valued mapping $\rho : \X\to\R$ is a \emph{risk measure} if it is:

\begin{enumerate}
\item[(i)] quasiconvex:  for all $x,y\in\X$ and $\lambda\in[0,1]$, $\rho(\lambda x+(1-\lambda)y) \leq \max\{\rho(x),\rho(y)  \}$
\item[(ii)] monotone: for all $x,y\in\X$, $x\geq y \implies \rho(x) \leq \rho(y)$
\end{enumerate}
\end{defi}

The following theorem states the bijective correspondence between risk orders and their representation via risk measures.

\begin{thm}[\citeasnoun{DrapeauKupper2013}]
Any numerical representation $\rho_{\hat\succsim} : \X\to\R$ of a risk order $\hat\succsim$ on $\X$ is a risk measure. Conversely, any risk measure $\rho:\X\to\R$ defines the risk order $\hat\succsim_\rho$ on $\X$ by
$$ x\hat\succsim_\rho y \iff \rho(x)\leq\rho(y) \ . $$
Risk orders and risk measures are bijectively equivalent in the sense that $\hat\succsim \ = \  \hat\succsim_{\rho_{\hat\succsim}}$ and $\rho_{\hat\succsim_\rho} = h \circ \rho$ for some increasing transformation $h:\R\to\R$.
\end{thm}

Note that, in the context of theories of choice, the risk measure $\rho_{\hat\succsim}$ corresponding to a given risk order $\hat\succsim$ is in fact the negative of the quasiconcave utility representation of the convex and monotonic total preorder $\hat\succsim$. In our setup of choice amongst risky positions $\LL^\infty(\Omega,\mathcal{F},\mathbb{P})$, a commonly used example of such a risk measure is the tail mean \cite{AcerbiTasche2002a,AcerbiTasche2002b}, defined by $\mathrm{TM}_\alpha = \E\left[ x \mid x>q_{\alpha}(x) \right]$, where $\alpha \in(0,1)$ is the confidence level and $q_\alpha(x) = \inf \{ x'\in\R : P(x\leq x') \geq\alpha \}$ is the lower $\alpha$-quantile of the random variable $x$.\footnote{This definition of tail mean holds only under the assumption of continuous distributions, that is for integrable $x$.}

Based on the previous discussion, a risk order exhibits preference for diversification (through the equivalent convexity axiom) if and only if the corresponding risk measure representing it is quasiconvex. This is a weakening of the general understanding of diversification within the theory of quantitative risk measurement, where diversification is encouraged when considering convex risk measures\footnote{A risk measure $\rho:\X\to\R$ is convex if for all $x,y\in\X$ and $\lambda\in[0,1]$, $\rho(\lambda x +(1-\lambda)y) \leq \lambda\rho(x) +(1-\lambda)\rho(y)$.} \cite{FollmerSchied2010,FollmerSchied2011} or, even more strongly, subadditive risk measures\footnote{A risk measure $\rho:\X\to\R$ is subadditive if for all $x,y\in\X$, $\rho(x+y) \leq \rho(x)+\rho(y)$.} \cite{Artzner1999}.

In mathematical finance, the passage from convexity to quasiconvexity is conceptually subtle but significant; see, for example, \citeasnoun{Cerreia2011}. While convexity is generally regarded as the mathematical formalization of the notion of diversification, it is in fact equivalent to the notion of quasiconvexity under a translation invariance assumption\footnote{A risk measure $\rho:\X\to\R$ is tranlsation invariant (or cash-additive) if for all $x\in\X$ and $m\in\R$, $\rho(x+m) = \rho(x)-m$.}. By considering the weaker notion of quasiconvex risk, one disentangles the diversification principle from the assumption of liquidity of the riskless asset -- an abstract simplification encapsulated through the translation invariance axiom. As we have seen above, the economic counterpart of quasiconvexity of risk measures is quasiconcavity of utility functions, which is equivalent to convexity of preferences.


\section{Notions of risk aversion}
\label{section:risk_aversion}

The concepts of diversification and risk aversion are closely intertwined. Informally, one might say that the goal behind introducing variety through diversification is the reduction of ``risk" or ``uncertainty", and so one might identify a diversifying decision maker with a risk averse one. We will later see that this is generally not the case.

In axiomatic choice theory, risk aversion is roughly the preference for a certain outcome with a possibly lower expected payoff over an uncertain outcome with equal or higher expected value.
In the economics literature, risk aversion is often exactly captured by the concavity of the utility function, and this is based on the underlying implicit framework of expected utility theory. In other models, however, risk aversion no longer goes along with a concave utility function, unless perhaps the very definition of risk aversion is reconsidered. In this section, when relating risk aversion to diversification or convexity of preferences, we look at intrinsic notions of risk aversion rather than model-dependent definitions. To this end, we use the three most frequently used definitions of weak, strong, and monotone risk aversion. In expected utility theory, all of these notions coincide and are characterized by the concavity of the utility function. We stress, however, the model-independency of the following definitions.

\subsection{Weak risk aversion}

The first, most common notion of risk aversion is based on the comparison between a random variable
and its expected value. A decision maker is \emph{weakly risk averse} if he always prefers the expected value of a random variable with certainty to the random variable itself.

\begin{defi}[Weak risk aversion]
The preference relation $\succsim$ on $\X$ is \emph{weakly risk averse} if $ \delta_{e(x)} \succsim x $ for every $x \in \X$, where $e(x)$ denotes the expected value of the random variable $x$. A decision maker is \emph{weakly risk seeking} if he always prefers any random variable to its expected value with certainty; formally if for all $x\in\X$, $ x \succsim \delta_{e(x)}$. A decision maker is \emph{weakly risk neutral} if he is always indifferent between any random variable and its expected value with certainty; formally if for all $x\in\X$, $ x \sim \delta_{e(x)}$.
\end{defi}

A straightforward characterization of weak risk aversion can be given in terms of the risk premium. Indeed, a decision maker is weakly risk averse if and only if the risk premium $\pi(x)$ associated to any $x\in\X$ is always nonnegative. Using this, one obtains a relation between decision makers ranking their level of risk aversion. Decision maker $D_1$ is said to be more risk averse than decision maker $D_2$ if and only if for every $x\in\X$, the risk premium $\pi(x)$ associated to $x$ is at least as great for $D_1$ as it is for $D_2$.

\subsection{Strong risk aversion}

The second notion of risk aversion is based on the definition of increasing risk of \citeasnoun{HadarRussell1969} and \citeasnoun{RothchildStiglitz1970} \citeaffixed{Landsberger1993}{see also} who define it in terms of the mean preserving spread.

\begin{defi}[Mean preserving spread]
For two random variables $x,y\in\X$, $y$ is a \emph{mean preserving spread} of $x$ if and only if $e(x)=e(y)$ and $x$ second-order stochastically dominates $y$, written as $x \succsim_{SSD} y$, that is if for any $C\in\R$,
$$ \int_{-\infty}^C F_x(c) dc \leq \int_{-\infty}^C F_y(c) dc. $$
\end{defi}

The mean preserving spread is intuitively a change from one probability distribution to another probability distribution, where the latter is formed by spreading out one or more portions of the probability density function or probability mass function of the former distribution while leaving the expected value unchanged. As such, the concept of mean preserving spreads provides a stochastic ordering of equal-mean choices according to their degree of risk. From the definition, we see that ordering choices by mean preserving spreads is a special case of ordering them by second-order stochastic dominance when the expected values coincide. Moreover, this ordering has the following properties.

\begin{lemma}
The stochastic ordering induced by the mean preserving spread on $x,y\in\X$ (i) depends only on the probability distributions of $x$ and $y$; and (ii) implies a non-decreasing variance, but a non-decreasing variance does not imply a mean preserving spread.
\end{lemma}

The notion of strong risk aversion can be viewed as aversion to any increase in risk, formalized next in terms of the mean preserving spread.

\begin{defi}[Strong risk aversion]
The preference relation $\succsim$ is \emph{strongly risk averse} if and only if for any $x,y\in\X$ such that $y$ is a mean preserving spread of $x$, $x \succsim y$. It is \emph{strongly risk seeking} if for any $x,y\in\X$ such that $y$ is a mean-preserving spread of $x$, $y\succsim x$. The preference relation $\succsim$ is \emph{strongly risk neutral} if for any $x,y\in\X$ such that $y$ is a mean-preserving spread of $x$, $x\sim y$.
\end{defi}

Preferences that are strongly risk averse (respectively strongly risk seeking) are also weakly risk averse (respectively weakly risk seeking). This is because for any $x\in\X$, $x$ is always a mean preserving spread of $\delta_{e(x)}$. Note also that strong and weak risk neutrality are equivalent, because weak risk neutrality implies that any $x$ is indifferent to $\delta_{e(x)}$ so that if $y$ is a mean preserving spread of $x$, they are both indifferent to $\delta_{e(x)}$.

\subsection{Monotone risk aversion}

The notion of strong risk aversion can be considered as too strong by some decision makers. This lead \citeasnoun{Quiggin1992} (see also \citeasnoun{Quiggin1991}) to define a new way for measuring increasing risk and, as a consequence, a new weaker notion of risk aversion, called \emph{monotone risk aversion}.

The definition of monotone risk aversion involves comonotonic random variables. Recall that two random variables $x,y\in\X$ are comonotonic if they yield the same ordering of the state space from best to worst. Clearly, every constant random variable is comonotonic with every other random variable. One can then define a measure of increasing risk of comonotonic random variables as follows.

\begin{defi}[Mean preserving monotone spread I]
Suppose $x$ and $y$ are two comonotonic random variables. Then $y$ is a \emph{mean preserving monotone spread} of $x$ if $e(x)=e(y)$ and $z=y-x$ is comonotonic with $x$ and $y$.
\end{defi}

This definition ensures that, since $x$ and $z$ are comonotonic, there is no hedging between $x$ and $z$, and thus that $y$ can be viewed as more risky than $x$. It can be extended to random variables that are not necessarily comonotonic, as follows.

\begin{defi}[Mean preserving monotone spread II]
For two random variables $x,y\in\X$, $y$ is a \emph{mean preserving monotone spread} of $x$ if there exists a random variable $\theta\in\X$ such that $y$ has the same probability distribution as $x+\theta$, where $e(\theta)=0$ and $x$ and $\theta$ are comonotonic.
\end{defi}

The notion of monotone risk aversion can be viewed as aversion to monotone increasing risk and is based on the definition of mean preserving monotone spread. We will later show that this concept of risk aversion is consistent with the rank-dependent expected utility theory of \citeasnoun{Quiggin1982}, one of the most well-known generalizations of expected utility theory, in which comonotonicity plays a fundamental part at the axiomatic level.

\begin{defi}[Monotone risk aversion]
The preference relation $\succsim$ on $\X$ is \emph{monotone risk averse} if for any $x,y\in\X$ where $y$ is a mean preserving monotone spread of $x$, $x\succsim y$. It is \emph{monotone risk seeking} if for any $x,y\in\X$ where $y$ is a mean preserving monotone spread of $x$, $y\succsim x$, and it is \emph{monotone risk neutral} if for any $x,y\in\X$ where $y$ is a mean preserving monotone spread of $x$, $x\sim y$.
\end{defi}

Finally, the following relationship between the three notions of weak, strong, and monotone risk aversion holds.

\begin{prop}[\citeasnoun{Cohen1995}]
Strong risk aversion implies monotone risk aversion; monotone risk aversion implies weak risk aversion; weak risk neutrality, strong risk neutrality, and monotone risk neutrality are identical.
\label{prop:risk_aversion}
\end{prop}


\section{Model-independent diversification preferences}
\label{section:model_ind}

We study the relationship between diversification preferences, convex preferences and risk aversion when preferences are not assumed to fit a specific choice theoretic model. As before, we simply assume that preferences are monotonic and continuous.

\subsection{Weak risk aversion and diversification}

Because the conventional definition of diversification is too strong to yield an equivalence to weak risk aversion when we move outside the assumptions of expected utility theory, the weaker concept of sure diversification was introduced. This weaker notion of diversification is indeed equivalent to weak risk aversion, independent of any model.

\begin{prop}[\citeasnoun{Chateauneuf2007}]
A monotonic and compact continuous\footnote{A preference relation $\succsim$ is \emph{compact continuous} if $x\succsim y$ whenever a bounded sequence $(x_n)_{n\in\N}$ converges in distribution to $x$ and $x_n\succsim y$ for each $n$. As noted by \citeasnoun{ChewMao1995}, many widely used examples of expected utility preferences are in fact compact continuous and not continuous when the corresponding utility function is discontinuous or unbounded. } preference relation $\succsim$ exhibits preference for sure diversification if and only if it is weakly risk averse.
\end{prop}

\subsection{Strong risk aversion and diversification}

In the space of probability distributions rather than random variables, \citeasnoun{Dekel1989}\footnote{The theoretical setup of \citeasnoun{Dekel1989} used to derive the results reviewed in this section is a very particular one, and we encourage the reader to read his article for the details.} shows that -- assuming no particular choice model -- preference for diversification is usually stronger that strong risk aversion. Indeed, he shows that preference for diversification implies risk aversion, but the converse is false.

\begin{prop}[\citeasnoun{Dekel1989}]
A strongly risk averse preference relation $\succsim$ does not necessarily exhibit preference for diversification.
\end{prop}

This means that preference for diversification is generally stronger than strong risk aversion. However, a complete characterization of strong risk aversion can be achieved through a weakening of preference for diversification obtained through the notion of strong diversification.

\begin{thm}[\citeasnoun{Chateauneuf2007}]
A monotonic and compact continuous preference relation $\succsim$ exhibits preference for strong diversification if and only if it respects second-order stochastic dominance.
\end{thm}

An immediate consequence of this result is that convexity of preferences, or equivalently preference for diversification, implies strong risk aversion. We point out that \citeasnoun{Dekel1989} proves the same result in the framework of probability distributions.

\begin{cor}
A preference relation $\succsim$ exhibiting preference for diversification is strongly risk averse.
\end{cor}


\section{Choice under risk I: expected utility theory}
\label{section:eut}

Since the publication of the seminal \emph{Theory of Games and Economic Behavior} of \citeasnoun{VNM1944}, expected utility theory (EUT) has dominated the analysis of decision-making under risk and has generally been accepted as a normative model of rational choice.\footnote{This is despite the evidence supporting alternative descriptive models showing that people's actual behavior deviates significantly from this normative model. See \citeasnoun{Stanovich2009} and \citeasnoun{HastieDawes2009} for a discussion.} A wide range of economic phenomena that previously lay beyond the scope of economic formalization were successfully modelled under expected utility theory.

This Section illustrates that, in the axiomatic setup of expected utility theory, (i) all notions of risk aversion coincide with concavity of utility, (ii) all notions of diversification coincide, and (iii) preference for diversification (that is convexity of preferences) is equivalent to risk aversion.

\subsection{Von Neumann--Morgenstern representation}

A utility representation of the preference relation $\succsim$ on $\X$ under EUT takes the form
$$ u(x) = \int u(c) \, dF_x(c), \quad x \in\X, \quad c\in\R, \quad u \colon \R \to \R.$$

Traditionally, the objects of choice in the von Neumann--Morgenstern setup are \emph{lotteries} rather than random variables, which are formalized via probability distributions. The agent's choice set is hence the set $\mathcal{M}$ of all probability measures on a separable metric space $(S,\mathcal{B})$, with $\mathcal{B}$ the $\sigma$-field of Borel sets. A utility representation $u$ of a preference relation $\succsim$ on $\mathcal{M}$ takes the form
$$u(\mu) = \int u(x) \mu(dx), \quad \mu \in \mathcal{M}, \quad u \colon S \to \R.$$
A \emph{compound lottery} that is represented by the distribution $\alpha \mu + (1-\alpha) \nu \in \mathcal{M}$, for $\alpha\in[0,1]$, gives $\mu$ with probability $\alpha$ and $\nu$ with probability $(1-\alpha)$, and so the probability of an outcome $x$ under the compound lottery is given by
$$ \left(\alpha\mu + (1-\alpha)\nu\right)(x) = \alpha\mu(x) + (1-\alpha)\nu(x). $$

The crucial additional axiom that identifies expected utility theory is the \emph{independence axiom}. It has also proven to be the most controversial. The independence axiom states that, when comparing the two compound lotteries $\alpha \mu + (1-\alpha)\xi$ and $\alpha \nu + (1-\alpha)\xi$, the decision maker should focus on the distinction between $\mu$ and $\nu$ and hold the same preference independently of both $\alpha$ and $\xi$. The key idea is that substituting $\xi$ for part of $\mu$ and part of $\nu$ should not change the initial preference ranking. The formal statement of the axiom follows.

\begin{enumerate}
\item[] \emph{Independence}: For all $ \mu, \nu, \xi \in \mathcal{M}$ and $\alpha \in [0,1]$, $\mu \succsim \nu \implies \alpha \mu + (1-\alpha)\xi \succsim \alpha \nu + (1-\alpha)\xi$.
\end{enumerate}

\begin{rem}
In this paper we adopted the classical setup in mathematical finance, which is the space of random variables. However, as mentioned above, the von Neumann--Morgenstern setup is the space of lotteries. In view of our discussion on preference for diversification, we refer to \citeasnoun[Proposition~3]{Dekel1989}, which states that convexity of preferences over the space of lotteries together with risk aversion implies preference for diversification over the space of random variables. This clarifies the link between convexity of preferences over the space of lotteries and convexity of preferences over the space of random variables.
\end{rem}

\subsection{Risk aversion in EUT}

 The power of the analysis of concepts of risk aversion and of the corresponding interpretation of increasing risk in terms of stochastic dominance contributed to a large degree to the success of EUT in studying problems relating to risk.
Indeed, in EUT, risk aversion corresponds to a simple condition on the utility function: within the class of preference relations which admit a von-Neumann-Morgenstern representation, a decision maker is characterized via his concave utility function.

\begin{prop}
Suppose the preference relation $\succsim$ satisfies expected utility theory and admits a von Neumann-Morgenstern utility representation $u$. Then: (i) $\succsim$ is weakly risk averse if and only if $u$ is concave; and (ii) $\succsim$ is strongly risk averse if and only if $u$ is concave.
\label{prop:eu_aversion}
\end{prop}

An immediate consequence of Proposition \ref{prop:eu_aversion} and Proposition \ref{prop:risk_aversion} is that, under the expected utility framework, all three notions of weak, strong, and monotone risk aversion coincide.

\begin{cor}
Under the assumptions of expected utility theory, the definitions of weak, strong, and monotone risk aversion are equivalent.
\end{cor}

In its essence, expected utility theory imposes restrictions on choice patterns. Indeed, it is impossible to be weakly risk averse without being strongly risk averse.
As a consequence, in EUT one simply speaks of ``risk aversion" without any need to specify the particular notion. Because it is characterized by concavity of utility, one can characterize the level of risk aversion through the curvature of the utility function. We recall the most commonly used such measure of risk aversion, introduced by \citeasnoun{Arrow1965} and \citeasnoun{Pratt1964}, and its characterization in the expected utility model.

\begin{defi}[Arrow-Pratt measure of risk aversion]
For an expected utility theory decision maker with utility function $u$, the \emph{Arrow-Pratt coefficient of absolute risk aversion} is defined for any outcome $c \in\R$ by
$$ A(c) = -\frac{u''(c)}{u'(c)}. $$
\end{defi}

\begin{prop}
Suppose that $u_1$ and $u_2$ are two strictly increasing and twice continuously differentiable functions on $\R$ representing expected utility preferences with corresponding Arrow-Pratt coefficients $A_1$ and $A_2$ and risk premiums $\pi_1$ and $\pi_2$, respectively. Then the following conditions are equivalent: (i) $A_1(c)\geq A_2(c)$ for all outcomes $c\in \R$; (ii) $u_1 = g \circ u_2$ for some strictly increasing concave function $g$; (iii) $\pi_1(x) \geq \pi_2(x)$ for all $x\in\X$.
\end{prop}

This essentially characterizes the relation of being more risk averse through the Arrow-Pratt coefficient.

\subsection{Diversification preferences in EUT}

Under the assumptions of expected utility theory, the two forms of sure and comonotone diversification are both represented by concavity of the utility index and consequently cannot be distinguished. Furthermore, they cannot be distinguished from the traditional notion of diversification, which corresponds to convexity of preferences.

\begin{prop}[\citeasnoun{Chateauneuf2002}]
Suppose $\succsim$ is a preference relation in the expected utility theory framework with utility index $u$. Then the following statements are equivalent:

\vspace{-10pt}\begin{enumerate}
\item[(i)] $\succsim$ exhibits preference for diversification
\item[(ii)] $\succsim$ exhibits preference for sure diversification
\item[(iii)] $\succsim$ exhibits preference for comonotone diversification
\item[(iv)] $u$ is concave
\end{enumerate}
\end{prop}

Moreover, recall that the equivalence between diversification and risk aversion established in the expected utility framework does not hold in more general frameworks. In particular, the notion of strong diversification, that is  preference for diversification among two identically distributed assets, was shown to be equivalent to strong risk aversion. Under the assumptions of EUT, this simply means that strong diversification coincides with all forms of risk aversion and, therefore, with concavity of utility.

\begin{cor}
Suppose $\succsim$ is a preference relation in the expected utility theory framework with utility index $u$. Then the following statements are equivalent:
\begin{enumerate}
\item[(i)] $\succsim$ exhibits preference for strong diversification
\item[(ii)] $\succsim$ is risk averse
\item[(iii)] $u$ is concave
\end{enumerate}
\end{cor}

In summary, all notions of risk aversion (weak, strong, monotone) and of diversification (sure, strong, comonotone, and convex preferences) introduced in this article coincide with concavity of utility in the framework of expected utility theory.


As an illustration of a concrete portfolio choice implication of this correspondence, consider the problem of investing in one risk-free asset with return $r$ and one risky asset paying a random return $z$. Under a full investment assumption, suppose the decision maker invests $w$ in the risky asset and the remaining $1-w$ in the safe asset, implying an overall portfolio return of $wz+(1-w)r$. Maximizing utility $u$ formally means
$$ \max_w \int u (wz+(1-w)r) dF(z) \ ,$$
where $F$ is the cumulative distribution function (cdf) of the risky asset.
An investor who is risk neutral will only care about the expected return, and hence will put all of the available wealth into the asset with the highest expected return. If the investor is risk averse, the utility is concave, and hence the second order condition of the maximization problem completely characterizes the solution. More importantly, this means that if the risky asset has a postitve rate of return above the risk-free rate (that is $z>r$), the risk averse investor will still choose to invest some amount $w>0$ in the risky asset. In summary, under the diversification preferences of expected utility maximizers, all risk averse decision makers will still want to take on some amount of risk at positive returns.

In his so-called local risk-neutrality theorem, \citeasnoun{Arrow1965} formally derives the expected utility portfolio choice pattern. Indeed, an economic agent who must allocate wealth between a safe and a risky asset will invest in an asset if and only if the expected value of the asset exceeds the price. The amount of the asset bought depends on the agent's attitude towards risk. Conversely, if the expected value is lower than the price of the asset, the agent will want to sell the asset short. Consequently, an investor's demand for an asset should be positive below a certain price, negative above that same price, and zero at exactly that price. In case there are many risky assets, this price will not necessarily be the expected value. Arrow's result holds in the absence of transaction costs whenever it is possible to buy small quantities of an asset.

Analyzing how changes in risk preferences affect optimal portfolio choice, one can use the comparative risk aversion measure of \citeasnoun{Arrow1965} and \citeasnoun{Pratt1964} to show the following intuitive result. If an expected utility investor with utility $u$ is more risk-averse than a second expected utility investor with utility $v$, then in the portfolio choice problem, the former investor will optimally invest less in the risky asset than the latter for any initial level of wealth. However, \citeasnoun{Ross1981} has pointed out that this result depends crucially on the fact that the less risky asset has a certain return. If both assets are risky, with one less risky than the other, a much stronger condition is needed to guarantee that the less risk averse investor will invest more in the riskier asset than the more risk averse investor for equal levels of wealth.


\section{Choice under risk II: rank-dependent expected utility theory}
\label{section:rdeu}

Rank-dependent expected utility theory (RDEU) is a generalization of expected utility theory accommodating the observation that economic agents both purchase lottery tickets (implying risk-seeking preferences) and insure against losses (implying risk aversion). In particular, RDEU explains the behaviour observed in the Allais paradox by weakening the independence axiom. RDEU was first axiomatized by \citeasnoun{Quiggin1982} as \emph{anticipated utility theory}, and was further studied by \citeasnoun{Yaari1987}, \citeasnoun{Segal1989}, \citeasnoun{Allais1987}, and \citeasnoun{Wakker1990}, amongst others.

After reviewing the basics of RDEU, we show that the equivalence between weak and strong risk aversion does not carry over from the expected utility model; see \citeasnoun{Machina1982b} and \citeasnoun{Machina1983}. Similarly, the correspondence of preference for diversification and risk aversion fails in the RDEU framework.

\subsection{Overview of RDEU}

The development of rank-dependent expected utility theory was motivated by the idea that equally probable events should not necessarily receive the same decision weights. Such a probability weighting scheme is meant to incorporate the apparent feature of overweighting of low probability events with extreme consequences that has been observed in violations of EUT models.

The RDEU model has a simple formalization in which outcomes are transformed by a von-Neumann-Morgenstern utility function and probabilities are transformed by a weighting function. The utility function being the same as in EUT implies that standard tools of analysis developed for EUT may be applied, with some modifications, to the RDEU framework. The probability weighting scheme first arranges states of the world so that the outcomes they yield are ordered from worst to best, then gives each state a weight that depends on its ranking as well as its probability. These ideas are formalized as follows.

\begin{defi}[Preferences under rank-dependent expected utility theory]
A decision maker satisfies rank-dependent expected utility (RDEU) theory if and only if his preference relation $\succsim$ can be represented by a real-valued function $V$ such that for every $x,y\in\X$,
$$ x\succsim y \iff V_{f,u}(x) \geq V_{f,u}(y) $$
where
$V_{f,u}$ is defined for every $z\in\X$ by
$$ V_{f,u}(z) = \int_{-\infty}^0 \left( f(\prob{u(z)>t}) -1 \right) dt + \int_0^\infty f\left( \prob{u(z)>t}\right) dt \ , $$
where $u:\R\to\R$, the utility function representing $\succsim$, is assumed to be continuous, strictly increasing  and unique up to positive affine transformations, and $f:[0,1]\to[0,1]$ is a unique, continuous and strictly increasing function satisfying $f(0)=0$ and $f(1)=1$.

When $\X$ has a finite number of outcomes $x_1\leq x_2\leq \cdots\leq x_n$, this representation reduces to
$$ V_{f,u}(z) = u(x_1) + \sum_{i=2}^n f\left( \sum_{j=i}^n p_j \right) (u(x_i) - u(x_{i-1})) \ . $$
\end{defi}

Preferences under RDEU are therefore characterized by the functions $u$ and $f$; the utility function $u$ is interpreted as the utility level under certainty, and the transformation function $f$ is interpreted as the perception of probabilities.

Note that in the case that $f(p)=p$ for all $p\in[0,1]$, RDEU reduces to expected utility theory. Consider on the other hand the case $f(p)\leq p$, which means that the decison maker's perception of probability is less than the actual probability. For finite $\X$, this condition implies that the decision maker, having at least utility $u(x_1)$, systematically underweights the additional utilities $u(x_i)-u(x_{i-1})$. Such a decision maker is referred to as being \emph{(weakly) pessimistic}.

\begin{defi}
A RDEU preference relation $\succsim$ is \emph{weakly pessimistic} if and only if $f(p)\leq p$ for all $p\in[0,1]$, and \emph{weakly optimistic} if and only if $f(p)\geq p$ for all $p\in[0,1]$.
\end{defi}

\subsection{Risk aversion in RDEU}

RDEU models suggest an approach to risk aversion that differs from EUT. By definition, RDEU theory can be viewed as embodying a fundamental distinction between attitudes to outcomes and attitudes to probabilities. Risk aversion within the RDEU framework should then encompass two different phenomena. The first is the standard notion of risk aversion within EUT associated with preferences over outcomes in terms of declining utility of wealth. The second relates to preferences over probabilities, that is to the transformation function $f$. We next review characterizations of both the utility function $u$ and the probability weighting function $f$ in terms of the various notions of risk aversion.\footnote{For a more complete review of the notions of risk aversion within the theory of choice under risk, we refer the reader to \citeasnoun{Cohen1995}.}

We begin with weak risk aversion. \citeasnoun{ChateauneufCohen1994} give necessary and sufficient conditions under which preferences are weakly risk averse. We only mention one particular case in which their conditions  are both necessary and sufficient.

\begin{prop}[\citeasnoun{ChateauneufCohen1994}]
A RDEU decision maker whose utility function $u$ is concave and differentiable is weakly risk averse if and only if he is weakly pessimistic.
\end{prop}

Strong risk aversion in rank-dependent utility theory has been characterized by \citeasnoun{ChewKarniSafra1987} as follows.

\begin{prop}[\citeasnoun{ChewKarniSafra1987}]
A RDEU preference relation $\succsim$ is strongly risk averse if and only if its utility $u$ is concave and probability weighting $f$ is convex.
\end{prop}

This is a rather strong characterization, since under RDEU, a decision maker cannot be strongly risk averse without having concave utility. On the other hand, under the dual theory of \citeasnoun{Yaari1987}, strong risk aversion corresponds to the convexity of $f$ (see \citeasnoun{Yaari1987}).

Finally, monotone risk aversion under the RDEU framework has been characterized by \citeasnoun{Quiggin1992}.

\begin{prop}[\citeasnoun{Quiggin1992}]
(i) A RDEU decision maker who is monotone risk averse and whose utility $u$ is concave is weakly pessimistic. (ii) A RDEU decision maker who is weakly pessimistic and has concave utility $u$ is monotone risk averse.
\end{prop}

Moreover, one obtains a characterization of the mean-preserving monotone spread.

\begin{prop}[\citeasnoun{Quiggin1992}]
For $x,y\in\X$ for which $e(x)=e(y)$, $y$ is a monotone mean-preserving spread of $x$ if and only if for the preference relation $\succsim$ under RDEU, $x\succsim y$.
\end{prop}

\subsection{Diversification preferences in RDEU}

In expected utility theory, risk averse decision makers will always prefer a diversified portfolio over a concentrated one. A similar result holds for RDEU decision makers who are \emph{strongly} risk averse in the sense of second-order stochastic dominance.

\begin{prop}[\citeasnoun{Quiggin1993}]
A RDEU decision maker exhibits preference for diversification if and only if he is strongly risk averse, that is  if and only if the utility $u$ is concave and probability weighting $f$ is convex.
\end{prop}

The weaker notion of sure diversification, where the decision maker prefers diversification if he can attain certainty by a convex combination of choices (in addition to being indifferent between these choices), is actually equivalent to that of weak risk aversion in the RDEU framework.

\begin{prop}[\citeasnoun{Chateauneuf2007}]
A RDEU decision maker exhibits preference for sure diversification if and only if he is weakly risk averse.
\end{prop}

Assuming additionally concavity (and differentiability of utility), we can immediately relate sure diversification and the notion of weak pessimism under RDEU as follows.

\begin{cor}
A RDEU decision maker with differentiable and concave utility $u$ exhibits preference for sure diversification if and only if he is weakly pessimistic.
\end{cor}

Finally, consider the case of comonotonic random variables and the associated notion of comonotone diversification, which essentially restricts preference for diversification to comonotonic choices. Given comonotonic random variables, one can show that a RDEU decision maker will in fact be indifferent regarding diversification across such comonotonic prospects. This was proven by \citeasnoun{Roell1987}.

\begin{rem}[Convexity of preferences in RDEU]
\citeasnoun{Quiggin1993} discusses the relationship between diversification as a linear mixture of random variables and the concepts of convexity and concavity of preferences. Recall that we identify convex preferences with the traditional definition of diversification. However, in RDEU theory, where preference for diversification over outcome mixtures arises without convexity, this analogy is misleading, even though the correspondence between strong risk aversion and diversification carries over to this model. We refer the reader to \citeasnoun{Quiggin1993} for a detailed discussion and further references on this topic.
\end{rem}

\begin{rem}[Implications on portfolio choice]
Rank-dependent expected utility contains the following two components: a concave utility function and a reversed S-shaped probability distortion function. The first component captures the observation that individuals dislike a mean-preserving spread of the distribution of a random outcome. The second component captures the tendency to overweight tail events --- a principle that can explain
why people buy both insurance and lotteries. Indeed, \citeasnoun{Quiggin1991} points out that the behavior of an individual whose preferences are described by a RDEU functional with a concave outcome utility function and a (reversed) S-shaped probability weighting function will display risk aversion except when confronted with probability distributions that are skewed to the right.\footnote{A number of papers have studied portfolio theory, risk-sharing and insurance contracting in the RDEU framework; see \citeasnoun{Bernard2013} for a detailed review.}
\end{rem}


\section{Choice under uncertainty}
\label{section:ceu}

Models of choice under risk were originally formulated to be used with pre-specified or objective probabilities. One of their main limitations is that uncertainty is treated as objective risk.
Not all uncertainty, however, can be described by such an objective probability.
Following the fundamental works of \citeasnoun{Keynes1921}, \citeasnoun{Knight1921}, and \citeasnoun{Ramsey1926}, we now draw a distinction between uncertainty and risk --- \emph{risk} is used when the gamble in question has objectively agreed upon known odds, while \emph{uncertainty} roughly refers to situations where the odds are unknown.\footnote{Objective risk is typically available in games of chance, such as a series of coin flips where the probabilities are objectively known. In practice, the notion of risk also encompasses situations, in which reliable statistical information is available, and from which objective probabilities are inferred. Uncertainty, on the other hand, can arise in practice from situations of complete ignorance or when insufficient statistical data is available, for example.}


Notions of diversification and \emph{uncertainty} aversion, rather than \emph{risk} aversion, are hence discussed within models of expected utility under a (not necessarily additive) subjective probabilty measure, which seeks to distinguish between quantifiable \emph{risks} and unknown \emph{uncertainty}. Such choice theoretic models originated with the seminal works of \citeasnoun{Savage1954} and \citeasnoun{AnscombeAumann1963}. Our framework will be the axiomatic treatment of such models as developed by \citeasnoun{Schmeidler1989}; see also \citeasnoun{Gilboa1987}. These axiomatizations involve the use of the Choquet integral to derive the corresponding representation results for nonadditive probabilities --- we will hence refer to such models of uncertainty as \emph{Choquet expected utility (CEU)} models.

\subsection{Choquet expected utility theory}

Before formally setting up the axiomatic framework of \citeasnoun{Schmeidler1989}, we quickly review the basics of subjective expected utility theory under additive and nonadditive probability measures. In this subsection we depart from the theoretical setting of Section~\ref{sec:theoretical setup} and adopt the classical decisional theoretical setup where decision makers choose from a set of acts.

\subsubsection{Subjective expected utility}

In the standard model of (subjective) uncertainty pioneered by \citeasnoun{Savage1954}, the decision maker chooses from a set of acts. The formal model consists of a set of prizes or consequences $X=\R$ and a state space $S$ endowed with an algebra $\Sigma$. The set of acts $\A$ is the set of all finite measurable functions (the inverse of each interval is an element of $\Sigma$ from $\Omega$ to $\R$). Preference relations $\succsim$ are defined over acts in $\A$. A subjective expected utility representation of $\succsim$ is given by a subjective probability measure $\mathbb{P}$ on the states $S$ and a utility function $u\,:\,X\to\R$ on the consequences $X$ satisfying

$$ f\succsim g \iff \int_S u\left(f(s)\right) d\mathbb{P} \geq \int_S u\left(g(s)\right) d\mathbb{P}. $$

The expectation operation here is carried out with respect to a prior probability derived uniquely from the decision maker's preferences over acts.
Both \citeasnoun{Savage1954} and \citeasnoun{AnscombeAumann1963} provide axiomatizations of preferences leading to the criterion of maximization of subjective expected utility, making this representation hold, the latter being the simpler development, since a special structure is imposed on the set $X$ of consequences. The classic development of \citeasnoun{Savage1954} is more general yet more complex, a thorough review of which is beyond the scope of the present article. We do, however, briefly mention one key axiom of Savage's, known as the Sure-Thing Principle.

\begin{axiom}[Savage's Sure-Thing Principle]
Suppose $f,f',g,g'\in\A$ are four acts and $T\subseteq S$ is a subset of the state space such that $f(s) = f'(s)$ and $g(s) = g'(s)$ for all $s\in T$, and $f(s) = g(s)$ and $f'(s) = g'(s)$ for all $s\in T^C$. Then $f\succsim g$ if and only if $f'\succsim g'$.
\end{axiom}

This axiom can be interpreted in terms of preferences being separable --- if the decision maker prefers act $f$ to act $g$ for all possible states in $T^C$ (for example knowing a certain event will happen), and act $f$ is still preferred to act $g$ for all states in $T$ (for example if that certain event does not happen), then the decision maker should prefer act $f$ to act $g$ independent of the state. Using the example of an event occurring, this means that $f$ is preferred to $g$ despite having no knowledge of whether or not that certain event will happen. The axiom essentially states that outcomes which occur regardless of which actions are chosen, ``sure things", should not affect one's preferences.

\subsubsection{Schmeidler's axiomatization of subjective expected utility under nonadditive probabilities}

Schmeidler's axiomatization of choice under uncertainty without additivity formally separates the notion of individual perception of uncertainty from valuation of outcomes. Under the key axiom of comonotonic independence, preferences under uncertainty are characterized by means of a functional that turns out to be a Choquet integral.

Ellsberg's paradox\footnote{\citeasnoun{Ellsberg1961} proposed experiments where choices violate the postulates of subjective expected utility, more specifically the Sure-Thing Principle. The basic idea is that a decision maker will always choose a known probability of winning over an unknown probability of winning even if the known probability is low and the unknown probability could be a guarantee of winning. His paradox holds independent of the utility function and risk aversion characteristics of the decision maker, and implies a notion of \emph{uncertainty aversion}, which is an attitude of preference for known risks over unknown risks.} \cite{Ellsberg1961} was the main reason motivating the development of a more general theory of subjective probabilities, in which the probabilities need not be additive.

\begin{defi}[Nonadditive probability measure]
A real-valued set function $v\,:\,\Sigma \to [0,1]$ on an algebra $\Sigma$ of subsets of a set of states $S$ is a \emph{nonadditive probability measure} if it satisfies the normalization conditions $v(\emptyset) = 0$ and $v(S) = 1$, and the monotonicity condition, that is for all $A,B\in\Sigma$, if $A \subseteq B$, then $v(A)\leq v(B)$.
\end{defi}

We now formally set up Schmeidler's CEU model. The state space $S$ is endowed with an algebra $\Sigma$ of subsets of $\Omega$. The set of consequences is assumed to be the positive real line, i.e., $X=\R_+$. The set of acts is the set of $\LL$ nonnegative measurable functions on $S$. Preferences $\succsim$ of the CEU decision maker are defined over the set of acts $\LL$ and are assumed to be monotonic and continuous. In order to weaken the independence axiom, the notion of \emph{comonotonic independence} was introduced by \citeasnoun{Schmeidler1989}. Very roughly, it requires that the usual independence axiom holds only when hedging effects (in the sense of \citeasnoun{Wakker1990}) are absent:

\begin{axiom}[Comonotonic Independence]
For all pairwise comonotonic acts\footnote{Recall that two acts $f,g\in\LL$ are said to be \emph{comonotonic} if for no $s,t\in S$, $f(s) > f(t)$ and $g(s) < g(t)$.} $f,g,h\in\LL$ and for all $\alpha\in(0,1)$, $f\succsim g$ implies $\alpha\, f + (1-\alpha)\,h \succsim \alpha\,g +(1-\alpha)\,h$.
\end{axiom}

A nonadditive probability measure is referred to as a \emph{capacity} $v:\Sigma\to[0,1]$, and we assume throughout that there exists $A\in\Sigma$ such that $v(A)\in(0,1)$. The \emph{core} of a capacity $v$ is defined by
$$ \core(v) = \left\{ \pi : \Sigma\to\R_+ \mid \forall A\in\Sigma, \pi(A)\geq v(A) ; \ \pi(S) = 1 \right\} \ . $$

\citeasnoun{Schmeidler1989} proved that the preference relation $\succsim$ on $\LL$ satisfying comonotonic independence (and the usual monotonicity and continuity axioms) is represented through a Choquet integral with respect to a unique capacity $v$ rather than a unique additive probability measure.

\begin{defi}[Choquet integral]
The \emph{Choquet integral} (\citeasnoun{Choquet1954}) of a real-valued function $u\,:\,\LL\to\R$ on a set of states $S$ with respect to a capacity $v$ is defined by
$$ \int u(f(\cdot)) \ dv = \int_{-\infty}^0 \left(v(u(f)\geq t)-1\right) dt + \int_0^\infty v(u(f)\geq t) \ dt \ . $$
\end{defi}

Note that if the capacity $v$ is in fact an additive probability measure $p$, then the Choquet integral reduces to the mathematical expectation with respect to $p$.

\begin{thm}[Schmeidler's Representation Theorem (\citeasnoun{Schmeidler1989})]
Suppose that the preference relation $\succsim$ on $\LL$ satisfies the comonotonic independence, continuity, and monotonicity axioms. Then there exists a unique capacity $v$ on $\Sigma$ and an affine utility-on-wealth function $u\,:\,\R\to\R$ such that for all $f,g\in\LL$,
$$ f \succsim g \iff \int_S u(f(\cdot)) \ dv \geq \int_S u(g(\cdot)) \ dv \ . $$
Conversely, if there exist $u$ and $v$ as above with $u$ nonconstant, then the preference relation $\succsim$ they induce on $\LL$ satisfies the comonotonic independence, continuity, and monotonicity axioms. Finally, the function $u$ is unique up to positive linear transformations.
\end{thm}

We say that a function $V:\LL\to\R$ represents the preference relation $\succsim$ if for all acts $f,g\in\LL$, $f \succsim g \iff V(f) \geq V(g)$, where, under the axiomatization of CEU, $V(f)$ for $f\in\LL$, is given by the Choquet integral $\int_S u(f(\cdot)) \ dv$, where $u$ and $v$ satisfy the properties of the previous Theorem.

\subsection{Diversification preferences under uncertainty}


In the Choquet expected utility framework, preference for diversification is equivalent to the utility index being concave and capacity being convex. The following Theorem provides this characterization:

\begin{thm}[\citeasnoun{Chateauneuf2002}]
Assume that $u:\R_+\to\R$ is continuous, differentiable on $\R_{++}$ and strictly increasing, and let $V$ be the functional representing the preference relation $\succsim$ under the CEU model. Then the following statements are equivalent:
\vspace{-5pt}\begin{enumerate}
\item[(i)] $\succsim$ exhibits preference for diversification.
\item[(ii)] $V$ is concave.
\item[(iii)] $u$ is concave and $v$ is convex.
\end{enumerate}
\end{thm}

Recall that the weaker notion of sure diversification introduced by \citeasnoun{Chateauneuf2002} can be interpreted as an axiom of uncertainty aversion at large; if the decision maker can attain certainty by a convex combination of equally desirable random variables, then he prefers certainty to any of these random variables. Sure diversification hence embodies a notion of aversion to ambiguity (in the sense of imprecise probability) as well as a notion of aversion to risk.

In the context of Choquet expected utility, sure diversification does not have a full characterization in terms of the utility index $u$ and capacity $v$, as the following Theorem states.

\begin{thm}[\citeasnoun{Chateauneuf2002}]
Let $v$ be the capacity functional in the Choquet expected utility framework, and suppose that the utility index $u$ is continuous, strictly increasing and differentiable on $\R_{++}$. If the preference relation $\succsim$ exhibits preference for sure diversification, then $\core(v) \neq \emptyset$. On the other hand, if $\core(v)\neq \emptyset$ and $u$ is concave, then $\succsim$ exhibits preference for sure diversification.
\end{thm}

Characterization of sure diversification  under CEU is hence incomplete --- a decision maker with a non-concave utility index may or may not be a sure diversifier. In particular, \citeasnoun{Chateauneuf2002} give an example for a convex utility index and a capacity with an empty core which does not yield preference for sure diversification.\footnote{See Example 1 in \citeasnoun{Chateauneuf2002}.} However, the concavity of the utility index can be shown to be equivalent to a different form of diversification, namely that of comonotone diversification, which is a restriction of convexity of preferences to comonotone random variables.
Indeed, comonotone diversification is characterized completely by the concavity of the utility index in the CEU framework:

\begin{thm}[\citeasnoun{Chateauneuf2002}]
Let $v$ be the capacity functional in the Choquet expected utility  framework, and suppose that the utility index $u$ is continuous, strictly increasing and differentiable on $\R_{++}$. Then the preference relation $\succsim$ exhibits preference for comonotone diversification if and only if $u$ is concave.
\end{thm}

\begin{cor}[\citeasnoun{Chateauneuf2002}]
Let $v$ be the capacity functional in the Choquet expected utility framework, and suppose that the utility index $u$ is continuous, strictly increasing and differentiable on $\R_{++}$. Then the preference relation $\succsim$ exhibits preference for comonotone and sure diversification if and only if $u$ is concave and $\core(v)$ is non-empty.
\end{cor}

\subsection{Uncertainty averse preferences based on convexity}

\citeasnoun{Cerreira2011b} provide a characterization for a general class of preferences that are complete, transitive, convex and monotone, which they refer to as \emph{uncertainty averse preferences}. They establish a representation for uncertainty averse preferences in an Anscombe-Aumann setting which is general yet rich in structure.\footnote{Uncertainty averse preferences are a general class of preferences. Special cases that can be obtained by suitably specifying the uncertainty
aversion index G defined below include, amongst others, variational preferences and smooth ambiguity preferences. See \citeasnoun{Cerreira2011b} for more details.}

Let $\mathcal{F}$ be the set of all uncertain acts $f:S\to X$, where $S$ is the state space and $X$ is a convex outcome space, and let $\Delta$ be the set of all probability measures on $S$. Cerreira-Vioglio et al. show that a preference relation $\succsim$ is uncertainty averse (and satisfies additional technical conditions) if and only if there is a utility index $u:X\to\R$ and a quasi-convex function $G:u(X)\times\Delta\to(-\infty,\infty]$, increasing in the first variable, such that the preference functional
$$ V(f) = \min_{p\in\Delta} G \left( \int u(f) dp, p \right), \quad \forall f\in\mathcal{F} $$
represents $\succsim$, where both $u$ and $G$ are essentially unique.

In their representation, decision makers consider all possible probabilities $p$ and the associated
expected utilities $u(f)dp$ of act $f$. They then summarize all these evaluations by taking their minimum. The function $G$ can be interpreted as an index of uncertainty aversion; higher degrees of uncertainty aversion correspond to pointwise smaller indices $G$.

It is shown that the quasiconvexity of $G$ and the cautious attitude reflected by the minimum derive from the convexity of preferences. Uncertainty aversion is hence closely related to convexity of preferences. Under the formalization of Cerreiro-Vioglio et al., convexity reflects a basic negative attitude of decision makers towards the presence of uncertainty in their choices.

\section{Ambiguity aversion}
\label{section:ambiguity}

In the theory of choice under risk, the notion of risk aversion comes from a situation where a probability can be assigned to each possible outcome of a situation. In models of choice under uncertainty, however,  probabilities of outcomes are unknown, and thus the idea of a decision maker being risk averse makes little sense when such risks cannot be quantified objectively. Under uncertainty, the phenomenon of \emph{ambiguity aversion} roughly captures the preference for known risks over unknown risks. Whereas risk aversion is defined by the preference between a risky alternative and its expected value, an ambiguity averse individual would rather choose an alternative where the probability distribution of the outcomes is known over one where the probabilities are unknown.

The use of the term \emph{ambiguity} to describe a particular type of uncertainty is due to Daniel Ellsberg \cite{Ellsberg1961}.\footnote{``The nature of one's information concerning the relative likelihood of events... a quality depending on the amount, type, reliability and `unanimity' of information, and giving rise to one's degree of `confidence' in an estimation of relative likelihoods." \cite{Ellsberg1961}} As his primary examples, Ellsberg offered two experimental decision problems, which continue to be the primary motivating factors of research on ambiguity aversion.\footnote{They are referred to as the \emph{Two-Urn Paradox} and the \emph{Three-Color Paradox} -- see \citeasnoun{Ellsberg1961}.}
Unlike the economic concept of risk aversion, but similar to the notion of diversification, there is not unanimous agreement on what ambiguity aversion, also often referred to as uncertainty aversion, formally is. However several models and definitions have been proposed. We recall the most prominent such definitions and link them to the notion of diversification and discuss the implication of ambiguity aversion on portfolio choice.

\subsection{Schmeidler's uncertainty aversion}

\citeasnoun{Schmeidler1989} introduced a similar notion of aversion towards the unknown and called it \emph{uncertainty aversion}.
Uncertainty aversion roughly describes an attitude of preference for known risks over unknown risks. Formally, it is defined through convexity of preferences:

\begin{defi}[Schmeidler's uncertainty aversion]
A preference relation $\succsim$ on $\LL$ is said to exhibit \emph{uncertainty aversion} if for any two acts $f,g\in\LL$ and any $\alpha\in[0,1]$, we have
$$ f\succsim g \implies \alpha\, f + (1-\alpha)\,g \succsim g \ . $$
\end{defi}

To present an intuition for this definition, \citeasnoun{Schmeidler1989} explains that uncertainty aversion encapsulates the idea that ``\emph{smoothing or averaging utility distributions makes the decision maker better off}." This quote once again represents the general notion of diversification as discussed throughout this paper, namely that ``if $f$ and $g$ are preferred to $h$, so is any convex mixture $\lambda f + (1 - \lambda) g$ with $\lambda \in (0, 1)$."

Recall that in expected utility theory, risk aversion, convexity of preferences, and concavity of utility representations all coincide.
On the other hand, in Choquet expected utility models, convexity of preferences, that is aversion towards uncertainty, can be shown to be equivalent to convexity of capacity.

\begin{thm}[\citeasnoun{Schmeidler1989}]
A preference relation $\succsim$ exhibits uncertainty aversion if and only if the capacity $v$ is convex.
\end{thm}

The uncertainty aversion notion of \citeasnoun{Schmeidler1989} represents the first attempt to
formalize the notion that individuals dislike ambiguity. The intuition is that, by mixing two acts, the individual may be able to hedge against variation in utilities, much like, by forming a portfolio consisting of two or more assets, one can hedge against variation in monetary payoffs.

\begin{rem}[Alternative notions of ambiguity aversion]
Other attempts to characterize a dislike for ambiguity have been proposed in the
literature.
\citeasnoun{Epstein1999} proposed an alternative definition of uncertainty aversion that is more suited to applications in a Savage domain. His motivation is the weak connection between convexity of capacity and behavior that is intuitively uncertainty averse (see \citeasnoun{Epstein1999} and \citeasnoun{Zhang2002} for examples).
Yet another approach was proposed by \citeasnoun{Ghirardato2002}.
They consider both the Savage setting, with acts mapping to prizes, and the horse-roulette act framework, with acts mapping to objective probability distributions over prizes, and restrict their attention to preferences that admit a Choquet Expected Utility representation on binary acts, but are otherwise arbitrary.
For a comprehensive review of the notion of ambiguity aversion, we refer the reader to \citeasnoun{Machina2014}.
\end{rem}

\subsection{Ambiguity aversion and portfolio choice}


\citeasnoun{Keynes1921} was perhaps the first economist to grasp the full significance of uncertainty for economic analysis and portfolio choice. Whereas conventional models of choice under risk promote diversification,  Keynes expressed the view that one should allocate wealth in the few stocks about which one feels most favorably.\footnote{
{\it ``As time goes on I get more and more convinced that the right method in investment is to put fairly large sums into enterprises which one thinks one knows something about
and in the management of which one thoroughly believes. It is a mistake to think that
one limits one's risk by spreading too much between enterprises about which one knows
little and has no reason for special confidence. [...] One's knowledge and experience are definitely limited and there are seldom more than two or three enterprises at any given
time in which I personally feel myself entitled to put full confidence."}
See \cite{Keynes1983}.}
Indeed, even though Markowitz's idea of diversification has been accepted as one of the most fundamental tenets of modern financial economics, empirical evidence suggests that investors do not hold diversified portfolios but rather invest heavily in only a few assets and typically those with which they are \emph{familiar}.\footnote{See \cite{DeGiorgiMahmoud2014b} for a review of empirical evidence suggesting underdiversification.} Aversion to ambiguity essentially captures tilting away from the unknown and preference for the familiar. \citeasnoun{Boyle2012} show that if investors are familiar about a particular asset, they tilt their portfolio toward that asset, while continuing to invest in other assets; that is, there is both concentration in the more familiar asset and diversification in other assets. If investors are familiar about a particular asset and sufficiently ambiguous about all other assets, then they will hold only the familiar asset, as Keynes would have advocated. Moreover, if investors are sufficiently ambiguous about all risky assets, then they will not participate at all in the equity market. Their model also shows that when the level of average ambiguity across all assets is low, then the relative weight in the familiar asset decreases as its volatility increases; but the reverse is true when the level of average ambiguity is high. An increase in correlation between familiar assets and the rest of
the market leads to a reduction in the investment in the market. And, even when the number
of assets available for investment is very large, investors continue to hold familiar assets.

\citeasnoun{Dimmock2014} provide empirical evidence that ambiguity aversion
relates to five household portfolio choice puzzles: non-participation in equity markets, low
portfolio fractions allocated to equity, home-bias, own-company stock ownership, and portfolio
under-diversification. Consistent with the theory, ambiguity aversion is negatively associated with stock market participation, the fraction of financial assets allocated to stocks, and foreign stock ownership, but ambiguity aversion is positively related to owncompany stock ownership. Conditional on stock ownership, ambiguity aversion also helps to explain portfolio under-diversification.\footnote{A number of other research efforts empirically studying the effect of ambiguity aversion on portfolio choice reach the conclusion of under-diversification in some form, including the works of \citeasnoun{Uppal2003}, \citeasnoun{Maenhout2004}, \citeasnoun{Maenhout2006}, \citeasnoun{Garlappi2007}, \citeasnoun{Liu2010}, \citeasnoun{Campanale2011}, and \citeasnoun{Chen2014}.}

Under the definition of uncertainty aversion of \citeasnoun{Schmeidler1989}, \citeasnoun{Dow1992} derive the nonadditive analog of Arrow's local risk neutrality theorem. With a nonadditive probability distribution over returns on a risky asset, there is an interval of prices within which the economic agent has no position in the asset. At prices below the lower limit of this interval, the agent is willing to buy this asset. At prices above the upper end of the interval, the agent is willing to sell the asset short. The highest price at which the agent will buy the asset is the expected value of the asset under the nonadditive probability measure. The lowest price at which the agent sells the asset is the expected value of selling the asset short. This reservation price is larger than the other one if the beliefs reflect uncertainty aversion \`{a}-la-Schmeidler, that is, with a nonadditive probability measure, the expectation of a random variable is less than the negative of the expectation of the negative of the random variable. These two reservation prices, hence, depend only on the beliefs and aversion to uncertainty incorporated in the agent's prior, and not on attitudes or aversion towards risk.

\section{Some Concluding Behavioral Remarks}
\label{section:conclusion}

We have surveyed axiomatizations of the concept of diversification and their relationship to the related notions of risk aversion and convex preferences within different choice theoretic models. Our survey covered model-independent diversification preferences, preferences within models of choice under risk, including expected utility theory and the more general rank-dependent expected utility theory, as well as models of choice under uncertainty axiomatized via Choquet expected utility theory.

The traditional diversification paradigms of expected utility theory and Markowitz's Modern Portfolio Theory, which essentially encourage variety in investment over similarity, imply that individuals are rational and risk averse. However, experimental work in the decades after the emergence of the classical theories of \citeasnoun{VNM1944}  and \citeasnoun{Markowitz1952} has shown that economic agents in reality systematically violate the traditional rationality assumption when choosing among risky gambles. In response to this, there has been an explosion of work on so-called non-expected-utility theories, all of which attempt to better match the experimental evidence.  In particular, the last decade has witnessed the growing movement of behavioral economics, which questions the assumptions of rational choice theoretic models and seeks to incorporate insights from psychology, sociology and cognitive neuroscience into economic analysis. Behavioral economics has had some success in explaining how certain groups of investors behave, and, in particular, what kinds of portfolios they choose to hold and how they trade over time.


As an example, there is evidence suggesting that investors diversify their portfolio holdings much less than is recommended by normative models of rational choice. In particular, real-world diversification appears to be highly situational and context dependent; investors include a smaller number of assets in their portfolios than traditionally recommended; and some investors exhibit a pronounced home bias, which means that they hold only a modest amount of foreign securities. Moreover, the portfolio construction methodology recommended by Modern Portfolio Theory has some limitations. Consequently, alternative diversification paradigms have emerged in practice.

Because of the empirical work suggesting that investors' diversification behavior in reality deviates significantly from that implied by various rational models of choice, behavioral models assuming a specific form of irrationality have emerged. Behavioral economists turn to the extensive experimental evidence compiled by cognitive psychologists on the systematic biases that arise when people form beliefs.
By arguing that the violations of rational choice theory in practice are central to understanding a number of financial phenomena, new choice theoretic behavioral models have emerged.

\emph{Prospect Theory} is arguably the most prominent such behavioral theory. It states that agents make decisions based on the potential value of losses and gains rather than the final outcome, and that these losses and gains are evaluated using certain heuristics. The theory can be viewed as being descriptive, as it tries to model realistic observed and documented choices rather than optimal rational decisions. The theory is due to \citeasnoun{KahnemanTversky1979} and is largely viewed as a psychologically more accurate description of decision making compared to classical theories of rational choice. Since the original version of prospect theory gave rise to violations of first-order stochastic dominance, a revised version, called \emph{Cumulative Prospect Theory} (CPT), was introduced in \citeasnoun{KahnemanTversky1992}. CPT overcomes this problem by using a probability weighting function derived from rank-dependent expected utility theory.

We refer the reader to \citeasnoun{DeGiorgiMahmoud2014b} for a survey of the growing experimental and empirical evidence that is in conflict with how diversification preferences are traditionally viewed within classical models of choice, both under risk and under uncertainty, such as the one reviewed in this article. A particular focus is placed on Cumulative Prospect Theory. Unlike economic agents whose preferences are consistent with classical frameworks such as expected utility theory or Markowitz's Modern Portfolio Theory, decision makers whose preferences conform to Cumulative Prospect Theory do not select portfolios that are well-diversified. This lack of diversification essentially stems from inherent features of CPT, in particular those of framing and convex utility in the domain of losses.

\bibliographystyle{econometrica}
\bibliography{divreview_references,../refDB}

\end{document}